\documentclass[11pt,letterpaper,]{article}
\usepackage[T1]{fontenc}
\usepackage[utf8]{inputenc}
\usepackage{lmodern}
\usepackage{upgreek}
\usepackage{textcomp}
\usepackage{amsmath,amsfonts,amssymb,amsthm,mathrsfs}
\usepackage{latexsym}
\usepackage{authblk}
\usepackage{graphicx}
\usepackage[english]{babel}
\usepackage{xcolor}
\usepackage{soul}
\usepackage[normalem]{ulem}
\usepackage[outercaption]{sidecap}
\definecolor{apgreen}{rgb}{0.55, 0.71, 0.0}
\usepackage[top=2cm, bottom=2cm, left=2cm, right=2cm]{geometry}

\usepackage{abstract}
\addto{\captionsenglish}{}

\makeatletter
\renewcommand\@biblabel[1]{#1.} 
\makeatother

\usepackage[numbers,sort&compress]{natbib}  
\usepackage{setspace}
  
\emergencystretch=\maxdimen
\title{\bf \sffamily Intermittent migration can induce \\pulses of speciation in a two-island system}
\author[1,*]{Débora Princepe}
\author[2,*]{Simone Czarnobai}
\author[3]{Rodrigo A. Caetano}
\author[1,4]{Flavia M. D. Marquitti}
\author[1]{Marcus A. M. de Aguiar}
\author[3]{Sabrina B. L. Araujo}
\affil[1]{Instituto de F\'{i}sica `Gleb Wataghin', Universidade Estadual de Campinas, Campinas, Brazil}
\affil[2]{Programa de Pós Graduação em Ecologia e Conservação, Universidade Federal do Paraná, Curitiba, Brazil}
\affil[3]{Departamento de Física, Universidade Federal do Paraná, Curitiba, Brazil}
\affil[4]{ Instituto de Biologia, Universidade Estadual de Campinas, Campinas, Brazil}
\affil[*]{D.P. and S.C. should be considered joint first author}
\date{}

\begin{document}
\doublespacing 
\selectlanguage{english}
\maketitle

\begin{abstract}

\noindent Geographic barriers prevent migration between populations, thereby facilitating speciation through allopatry. However, these barriers can exhibit dynamic behavior in nature, promoting cycles of expansion and contraction of populations. Such oscillations cause temporal variations in migration that do not necessarily prevent speciation; on the contrary, they have been suggested as a driving force for diversification.
Here we present a study on a two-island neutral speciation model in scenarios with intermittent migration driven by sea-level fluctuations.
Mating is constrained to genetically compatible individuals inhabiting the same island, and offspring inherit nuclear genomes from both parents with recombination. We observe pulses of speciation that would not occur in strict isolation or continuous migration.
According to the seabed height, which modulates the duration of the isolation and connection periods, the maximum richness occurs at different times and in an ephemeral fashion. The expansion-contraction dynamics can accelerate diversification, but a long time in isolation can reduce the richness to one species per island, resembling patterns described by the taxon pulse hypothesis of diversification. Together with other studies, our results support the relevance of research on the impact of variable migration on diversification, suggested to be related to regions of high diversity.

\vspace{0.5cm}

\end{abstract}

\setlength{\parindent}{0.5cm}
\pagebreak



\section*{Introduction}

Migration -- the movement of organisms from one location to another -- has long been recognized as a key force driving the evolution of populations, influencing their genetic diversity and structure, and affecting the emergence of new species.
Nevertheless, the effects of migration on diversification can be complex and contrasting. 
On the one hand, a few migrants can facilitate speciation by increasing genetic diversity \citep{Lewis2023}, reinforcing reproductive barriers \citep{Matute2010}, and enabling the colonization of new areas \citep{Spurgin2014}. On the other hand, high levels of migration limit genetic differentiation by homogenizing previously isolated populations, which undermines diversification by isolation \citep{Claramunt2012} and opposes the effects of local selection \citep{Lenormand2002}.
Recent efforts to demonstrate speciation with gene flow have suggested, both theoretically \citep{Yamaguchi-2013_first,Princepe2022} and empirically \citep{Garant2007,Tigano2016}, that intermediate levels of migration can be optimal and allow divergence. This opposes the most accepted paradigm of allopatric speciation, which presupposes geographical isolation of populations for species diversification \citep{coyne_speciation_2004}. 
But just as strict isolation is rare in nature, continuous migration (recurring every generation) is also unrealistic, as assumed in most models and data analyses.

Several factors can cause temporal variability in migration rates at ecological and evolutionary time scales.
Behavioral and physiological demands, exemplified by seasonal relocation \citep{Everson2019}, predator-prey/host-parasite dynamics \citep{Hoberg2008}, or variation in population sizes \citep{Matthysen2005}, can affect the movement of organisms over individual and species lifetimes. Meanwhile, environmental variations, such as fluctuations in wind and oceanic currents \citep{White2010} or cyclic expansions and retractions of freshwater systems and tide pools \citep{Boizard2009,Tscha2017}, can intermittently act as geographical barriers that persist through evolutionary time. In this regard, sea-level fluctuations, triggered by geological and climatic events, are appraised as an essential mechanism in the evolution of biodiversity hotspots, as it promotes periodic isolation and re-connection of terrestrial and marine populations \citep{Ludt2015,Briggs2013}. For instance, eustatic fluctuations during the Pleistocene, with minima of up to 130m below the present level, allegedly had a role in diversification in regions of high diversity and endemism \citep{Hewitt2000,Guo2015,baggio2017}. The discontinuous gene flow resulting from intermittent migration is reflected in phylogenetic relationships and genetic and phenotypic patterns of species \citep{Cowie2008,Yesson2009,Fauvelot2020}. Nevertheless, separating this effect from simultaneous changes in environmental and ecological conditions remains a challenge.

The evolution of populations undergoing successive cycles of expansion and isolation due to intermittent barriers was previously discussed by Erwin, who proposed a model of radiation in bursts to explain the geographic history of carabids \citep{Erwin1979,Erwin1985}.
This model, known as the taxon pulse hypothesis, suggests that diversification results from periods of dispersal, when barriers breach, allowing previously isolated species to expand into new areas and contact each other, interspersed with episodes of vicariance, when geographic barriers reform and populations contract, occasionally favoring speciation \citep{Halas2005,Lim2008,Hoberg2010}.
This conjecture provides an explanation for a high diversity that could not be generated by purely allopatric speciation  \citep{Halas2005,Lim2008}. 
Reticulate evolution and areas with a mixture of endemic and widespread species are considered typical signatures of this process, creating a complex mosaic in time and space for species \citep{Schweizer2010,Bouchard2005,Esseghir2000,baggio2017}.
Yet, the intricate nature of these patterns poses challenges in their empirical inference and identification.

Several theoretical studies have explored migration that varies over time, pursuing to predict conditions for speciation \citep{Yamaguchi-2013_first,He2019} and provide a platform for inferring demographics and diversification rates from data \citep{Linck2019,Albert2017}. 
Models that integrate selection have shown that pulsed migration, i.e., episodic bursts of migration separated by periods in isolation, can present contrasting effects on local adaptation depending mostly on the selection regime (strength and type) \citep{Rice2009,Aguilee2011,Peniston2019,Aubree2023}. 
Generally, they agree that speciation with gene flow is feasible as long as migration does not overwhelm selection, which is more likely with a pulsed behavior.
Likewise, metacommunity models have demonstrated that landscape dynamics with fluctuating connectivity can surprisingly increase the number of coexisting species \citep{Palamara2023} and be beneficial to populations persistence (\cite{Reigada2015}, in this case, a pulsed dispersal model). Models considering mixed periods of contact and isolation tested against continuous migration, or isolation followed by prolonged contact, also corroborate the favorable impact of pulsedness on speciation and richness \citep{He2019,Linck2019}. Moreover, present genomic patterns are better fit under this model when using the input of geological records as proxy for migration times. These studies highlight the significance of temporal variability in migration rates on speciation processes and the dynamics of biodiversity \citep{Feder2019,He2019,Linck2019}.

Alongside, neutral models can provide insights into the effects of expansion-contraction dynamics alone, as well they have contributed to understanding diversity patterns and community assembly properties \citep{MacArthur1967,de2009global}. In particular, \cite{Manzo_Peliti_1994} and \cite{Princepe2022} explored a two-island neutral model where terrestrial individuals can migrate at a constant rate. This approach identified two mechanisms of speciation induced by migration at low and intermediary rates: founding populations, when a small influx of migrants incompatible with residents creates a sub-population that diverges, and induced sympatric speciation, when migrants mix with residents, and the resulting genetically diverse population eventually speciate \citep{Princepe2022}. 

Here we present a study built upon this prior research to investigate speciation in a scenario where isolation and connection occur cyclically, focusing on sea-level oscillations. The connectivity varies in time with the sea level: for a given seabed depth, the two islands are isolated when the sea level is above it. Alternately, when the sea level drops below this seabed, it forms a land bridge that connects the islands, and individuals can migrate with a given probability. In this analysis, we use the sea-level data of the past 800 thousand years, making explicit the dependence of the connection times on the seabed depth. Varying the migration probability during the connecting phase, we investigate how these parameters can affect species richness in the insular system. We also consider hypothetical periodic sea-level oscillations, where the isolation and connection have equal duration, to explore how different regimes induce speciation. Our results are discussed in light of existing models and empirical examples.

\section*{Methods}

We use individual-based simulations based on the models proposed by \cite{Manzo_Peliti_1994} and \cite{Princepe2022}: the individuals are equally distributed in two islands with a fixed carrying capacity and can migrate from one to another when the barrier separating them is lifted. 
Mating occurs only between genetically similar individuals inhabiting the same island. The population evolves due to recombination and mutation of their genomes, and speciation results from the breakdown of gene flow. 
We explore the effects of cyclic phases of isolation and connection imposed by sea-level fluctuations, adopting the historical records \citep{Spratt2016} to model the migration times. We also examine simulations with hypothetical periodic sea-level variation to unveil how each component of the system influences the speciation process.

\subsection*{Model }

\begin{figure}[!b]
\begin{center}
    \includegraphics[width=0.4\linewidth]{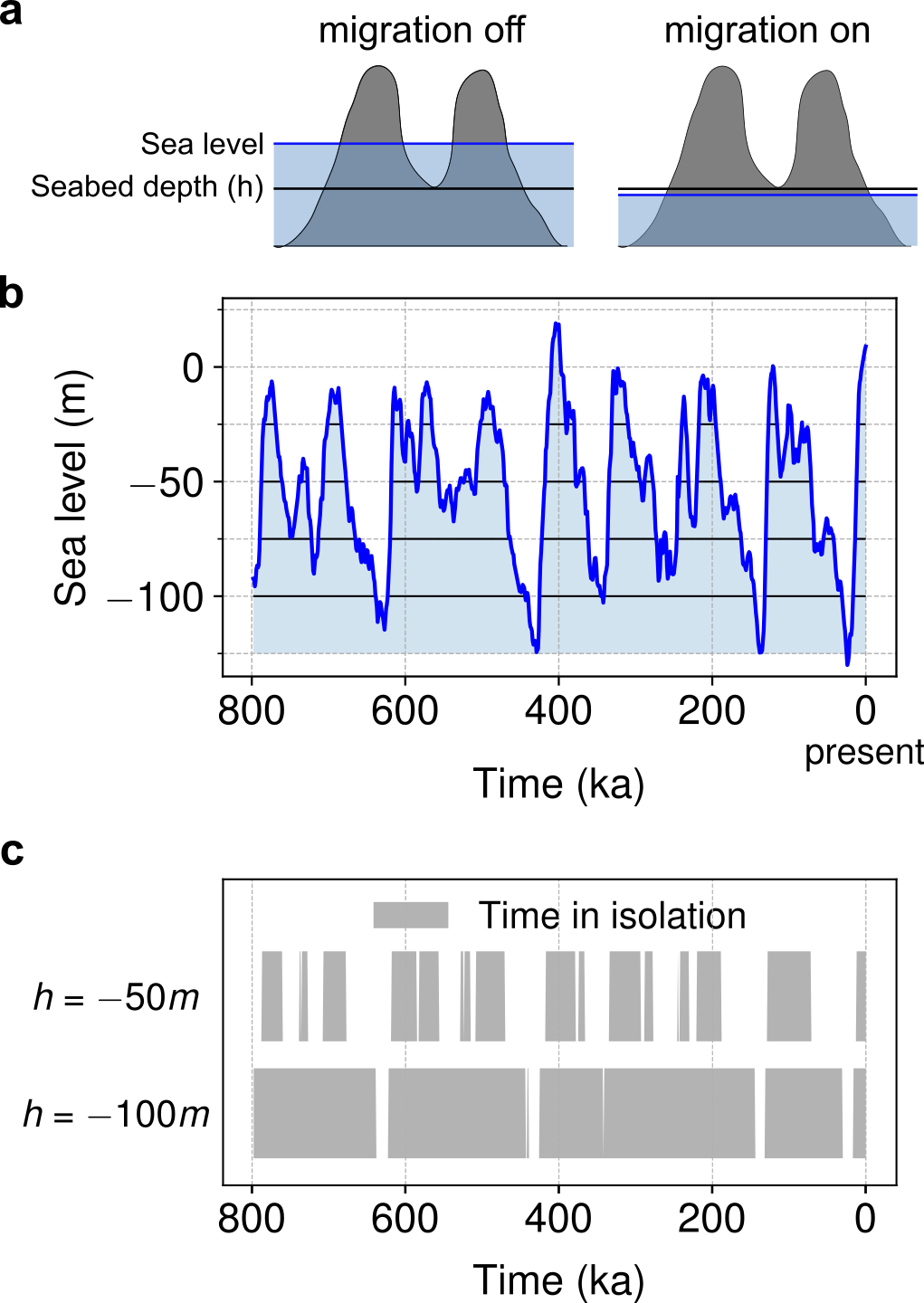}
\end{center}
\caption{Sea-level fluctuations as an intermittent barrier for migration. (a) We simulate a system of two islands where individuals can migrate from one island to another when the sea level is below the seabed. (b) Sea-level reconstruction over the last 800 thousand years (blue line, \cite{Spratt2016}). In each simulation, the seabed is fixed at a depth $h$, represented by horizontal lines (solid: time in isolation, dashed: connection). (c) Varying migration for $h=-50m$ and $h=-100m$: deeper seabeds result in longer periods of isolation. }
 \label{fig:fig1}
\end{figure}

Each individual is described by a biallelic genome, a chain of $B$ independent loci where each locus can assume the alleles 0 or 1 \citep{higgs_stochastic_1991,de2009global}. Individuals are hermaphroditic, and reproduction is sexual. The carrying capacity in each island $M$ (hereafter called population size) remains constant throughout the simulation, with small fluctuations allowed during the migration step \citep{Princepe2022}. The dynamic begins with $2M$ genetically identical individuals equally distributed in two islands and follows the steps at each iteration:

\textit{Migration:} if the islands are connected (when the sea level is below the seabed, see Fig. \ref{fig:fig1}a), each individual can migrate from one island to another with a probability $\epsilon$.

\textit{Reproduction:} individuals can reproduce with others on the same island. If migration is allowed, reproduction happens after the migration. $M$ descendants are born and replace the previous population in each island; thus, the generations do not overlap, and fluctuations in population size due to migration are compensated after reproduction. For each offspring, a first parent (P1) is chosen randomly, then the second parent (P2) is also randomly selected among all other individuals on the same island. To mate, the individuals must be compatible: their genetic distance (Hamming distance between the binary chains) must be less or equal to $G$ loci, where $G$ is a model parameter \citep{higgs_stochastic_1991,de2009global}. If they are incompatible, another individual P2 is drawn (with replacement) until a compatible partner for P1 is found. If no compatible individual is found after $M$ attempts, P1 is discarded, and a new random individual is selected (with replacement) as P1. The offspring's genome is a recombination of its parents' genome, inherited gene by gene with equal probability, followed by a  mutation probability $\mu$ per locus. 

\textit{Identifying species:} a species is defined by a group of individuals with gene flow established regardless of the island they inhabit, not only by direct genetic compatibility with others but also through intermediary individuals. The genetic distance between any pair of individuals belonging to different species is always higher than the threshold $G$, assuring reproductive isolation. 

Throughout the dynamics, species may disappear due to ecological drift, i.e., go extinct, or by hybridization of newly branched species. However, we have not differentiated between these two scenarios in the scope of this study.

In the limit where the genome size $B$ goes to infinite, sympatric speciation (in a single island) always occur if $G < G_0 = 2\mu M/(1+4\mu M)$ \citep{higgs_stochastic_1991,aguiar_speciation_2017}. However, speciation becomes rare as $B$ decreases below a critical value $B_c$, where $B_c$ depends on $\mu$, $M$ and $G$ \citep{aguiar_speciation_2017}. We shall use this property to choose the parameters for the simulations.

\subsection*{Sea-level fluctuations as an intermittent barrier} 

To investigate the impact of intermittent barriers on speciation, we used the sea-level data from \cite{Spratt2016}. This study is a compilation of several publications on sea-level reconstruction covering the last 800 thousand years (Fig. \ref{fig:fig1}b). Using this data, the sea level 5 thousand years ago is the reference for zero (the present level is estimated at $+8.96m$). 
We have scaled the time for computational feasibility: each model iteration was equivalent to 400 years; thus, we simulated the populations for 2000 iterations.

We varied the seabed depth, fixed at each simulation, to mimic different local geomorphological features. For example, if the sea level is at $-75m$, islands in a $-50m$ deep seabed region are connected, but where the seabed is $-100m$ deep, the islands remain isolated. Therefore, islands are isolated for longer periods on deeper seabeds (see Fig. \ref{fig:fig1}c).

\subsection*{Periodic sea-level fluctuations}

We also considered a simplified version for the sea-level oscillation with regular cycles, where the time in isolation is equal to the time with connection, given by the period $\tau$. In this case, the simulation starts with the islands isolated by $\tau$ years; then, they connect for $\tau$ years, and so on, adopting the same time scale as before (one model iteration equivalent to 400 years). Likewise, the migration only occurs during the connection phase, with a probability $\epsilon$ per individual. 

\subsection*{Simulation parameters} 

\cite{Princepe2022} investigated a similar model with continuous migration between two islands for a wide range of genome lengths and population sizes, demonstrating how different combinations of these parameters change the total and local species diversity. Based on this previous analysis, we chose a set of parameters such that the model dynamics in the absence of migration would result mainly in forming two species, one on each island, in agreement with strict allopatric speciation.
We verified that sympatric speciation under these conditions rarely occurred, resulting in an ensemble average of 1.1 species per island. Thus, comparing the present results with those obtained for continuous migration, we can isolate the role of the intermittent barrier in species formation.

We fixed the following parameters throughout the simulations: genome size ($B = 2,000$ loci), maximum genetic distance for mating ($G= 0.05B$), population size per island ($M = 200$), and mutation rate ($\mu = 0.001$ per locus per iteration). Populations evolved for $2000$ iterations, each corresponding to 400 years, which brings the mutation rate to $2.5\times10^{-6}$ per locus per year. 
We analyzed the variation of three parameters: (i) migration probability during the connection period, $\epsilon$, varied between 0 and 0.4 in steps of 0.02; (ii) seabed depth $h$, ranging from $-10m$ to $-100m$ in steps of $-5m$; and (iii) period $\tau$ for the simulations with regular cycles, varied between 20 and 400 thousand years in steps of 20 thousand years.
We ran 50 independent simulations for each combination of parameters and evaluated the ensemble-average species richness at specific points in the dynamics. Given that the system does not reach equilibrium, we also observed the trajectory of species richness for some configurations to understand the dynamic effect of the intermittent barrier.

\section*{Results}

\begin{figure}[ht]
\begin{center}
    \includegraphics[width=0.9\linewidth]{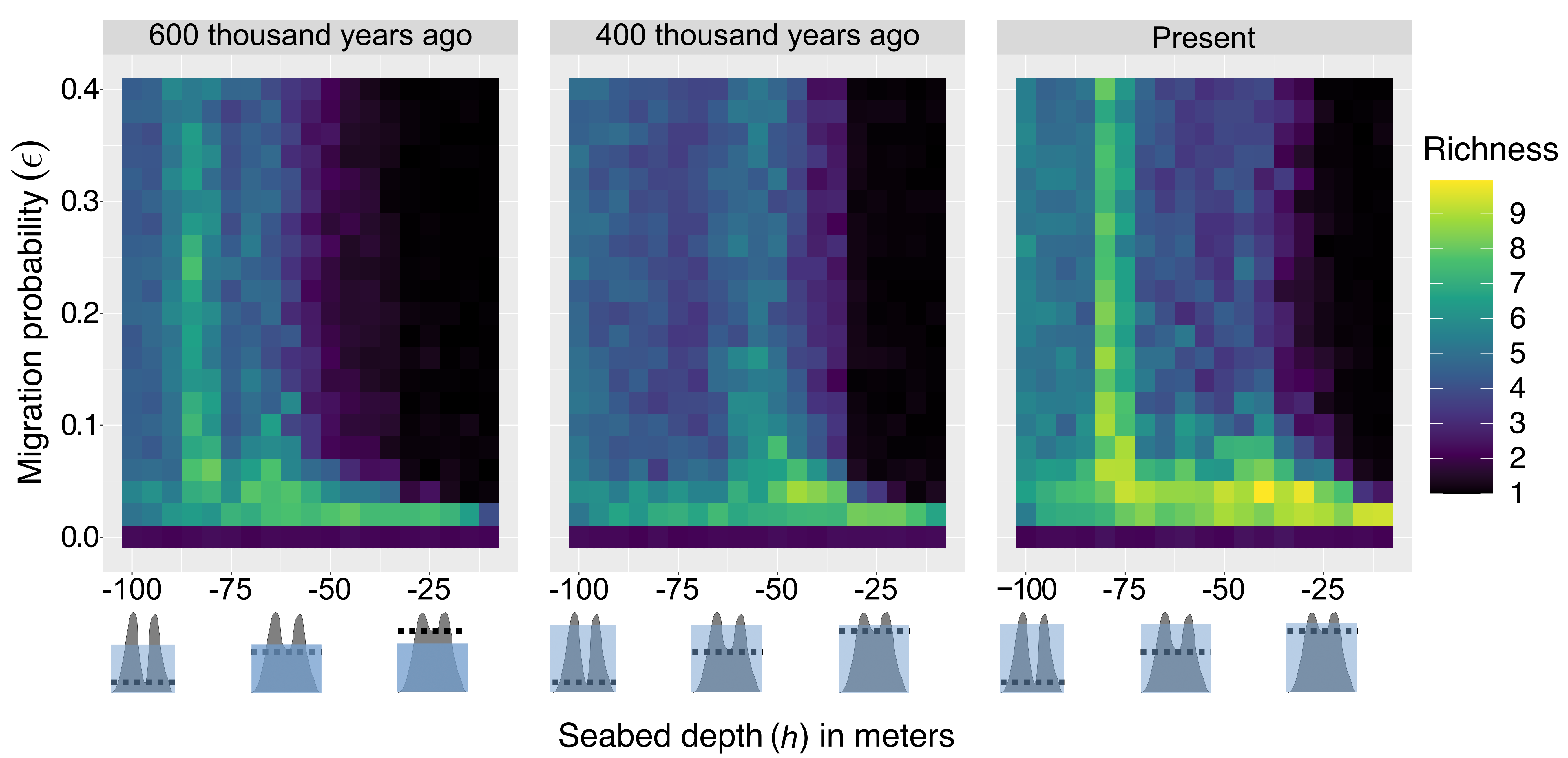}
\end{center}
\caption{Species richness as a function of migration probability ($\epsilon$) and seabed depth ($h$) for evolution under the sea-level fluctuations shown in Fig. 1b. Each plot shows the average species richness over 50 independent realizations measured at specific times. A low migration intensity ($\epsilon = 0.02$) resulted in a high number of species at any time, whereas the increase in richness varied over time for intermediate and large $\epsilon$, peaking at different seabed depths throughout the dynamics. }
 \label{fig:fig2}
\end{figure}

We first investigated how species richness depended on migration probability per individual ($\epsilon$) and seabed depth ($h$) throughout the dynamics employing the data on the sea-level fluctuations (\cite{Spratt2016}, Fig. \ref{fig:fig1}b). 
Figure \ref{fig:fig2} shows the average richness over 50 simulations at three different times: 600,000 years ago (left), 400,000 years ago (middle), and in the present (right), which respectively correspond to the measurement of the sea level at $-41.35$, $+18.54$, and $+8.96$ meters.

A low migration probability was sufficient to induce speciation that would otherwise be almost non-existent on completely isolated islands (Fig. \ref{fig:fig2}). With $\epsilon = 0.02$, high richness was sustained in time, with little variation along $h$ and slightly higher richness for shallow depths. However, for $\epsilon > 0.02$, the increase in richness depended on $h$ and the time of observation: speciation occurred in deep seabeds, where the isolation periods were longer, but was rare or absent in shallow ones. The boundary separating these two regions also changed over time: the dark region of the map, corresponding to a single species shared by both islands, gradually became limited to shallower depths. Furthermore, this increase in species richness was transitory, and peaks occurred at different moments for different seabed depths. For instance, depths around $h=-80m$ had peaks in richness in the oldest and most recent observations, while a transient increase occurred in the intermediate time for depths around $h=-50m$ (Fig. \ref{fig:fig2}).

\begin{figure}[!ht]
\begin{center}
    \includegraphics[width=0.9\linewidth]{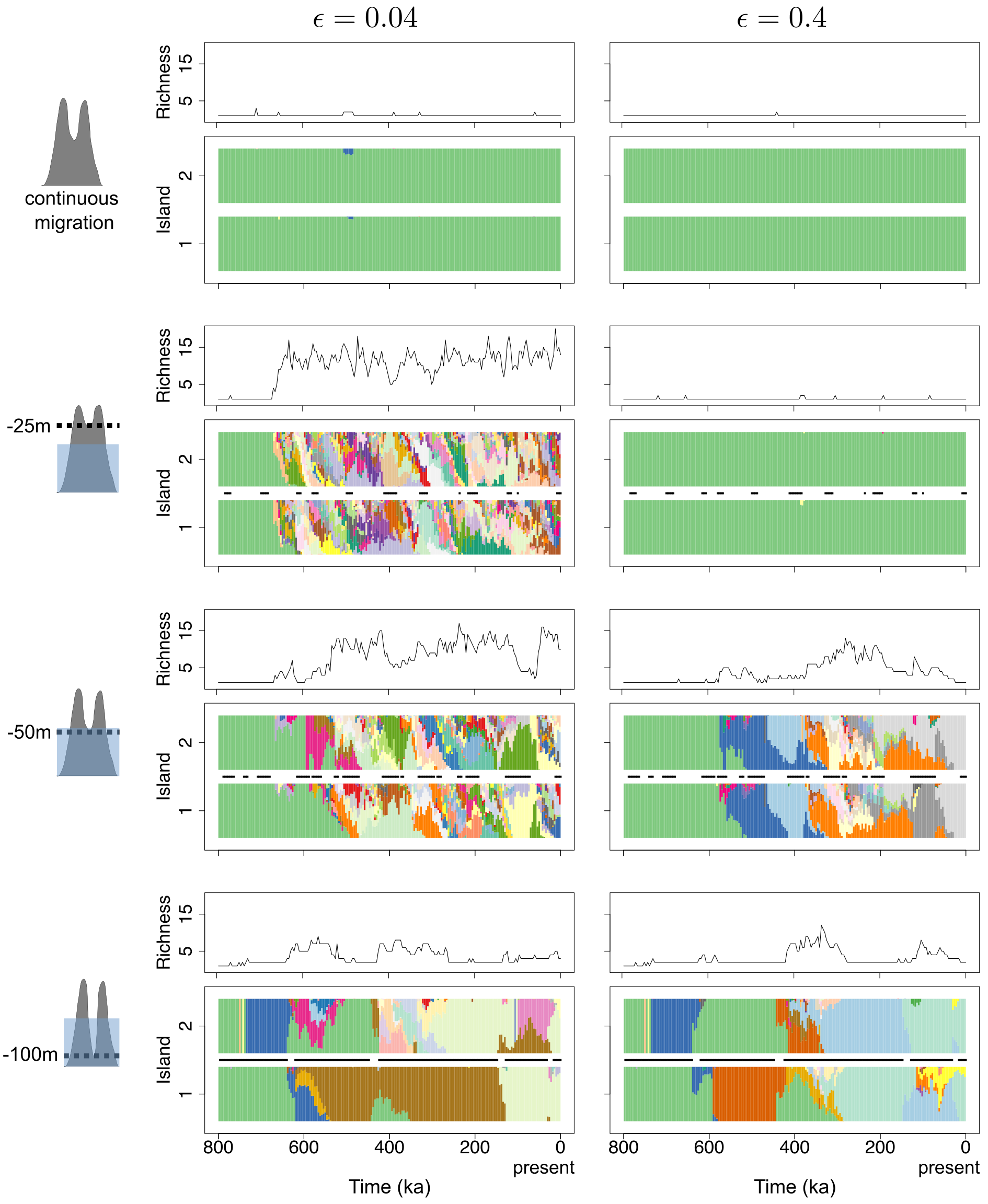}
\end{center}
\caption{Population evolution on each island in a single simulation for different seabed depths and migration probability $\epsilon=0.04$ and $\epsilon=0.4$. The first row considers the absence of a barrier, equivalent to continuous migration.
For each combination of parameters, the top graph shows the species richness throughout the simulation; the bottom graphs represent their distribution in each island, distinguished by colors and with vertical amplitude proportional to the species abundance. The black lines between the abundance plots indicate the time in isolation. }
 \label{fig:fig3}
\end{figure}

To understand this displacement of the richness peak over time, we looked at the temporal variation of species richness of single simulations for different seabed depths and $\epsilon = 0.04$ and $0.4$ (Fig. \ref{fig:fig3}). We also inspected the simulation without a barrier for comparison, where migration occurred continuously (Fig. \ref{fig:fig3}, first row). In this case, speciation was rare, and the islands had a single common species most of the time, regardless of the migration intensity.
With a shallow seabed ($h= -25m$, second row in Fig. \ref{fig:fig3}), sea-level fluctuations induced several cycles of short-term isolation and long-term connection. These short periods in isolation were enough to promote speciation under low migration probability ($\epsilon =  0.04$), but it got disrupted under high migration ($\epsilon = 0.4$). 
Indeed, a low migration probability was enough to promote speciation in all scenarios, while speciation under high migration only occurred for deeper seabeds, $h\leq-50m$. 
When the intermittence pattern of the barrier was the opposite, that is, cycles of long-term isolation periods followed by short-term connections ($h= -100m$, last row in Fig. \ref{fig:fig3}), migration had a dynamic effect on speciation: during the time the islands were connected, species from one island could colonize the other, and the subsequent isolation favored speciation, resembling a typical pattern described by the taxon pulse hypothesis of diversification \citep{Erwin1985,Halas2005}. 
However, the long time in isolation reduced diversity back to one species per island after the pulse, which is the expected equilibrium for fully isolated islands.

\begin{figure}[!ht]
\begin{center}
    \includegraphics[width=0.9\linewidth]{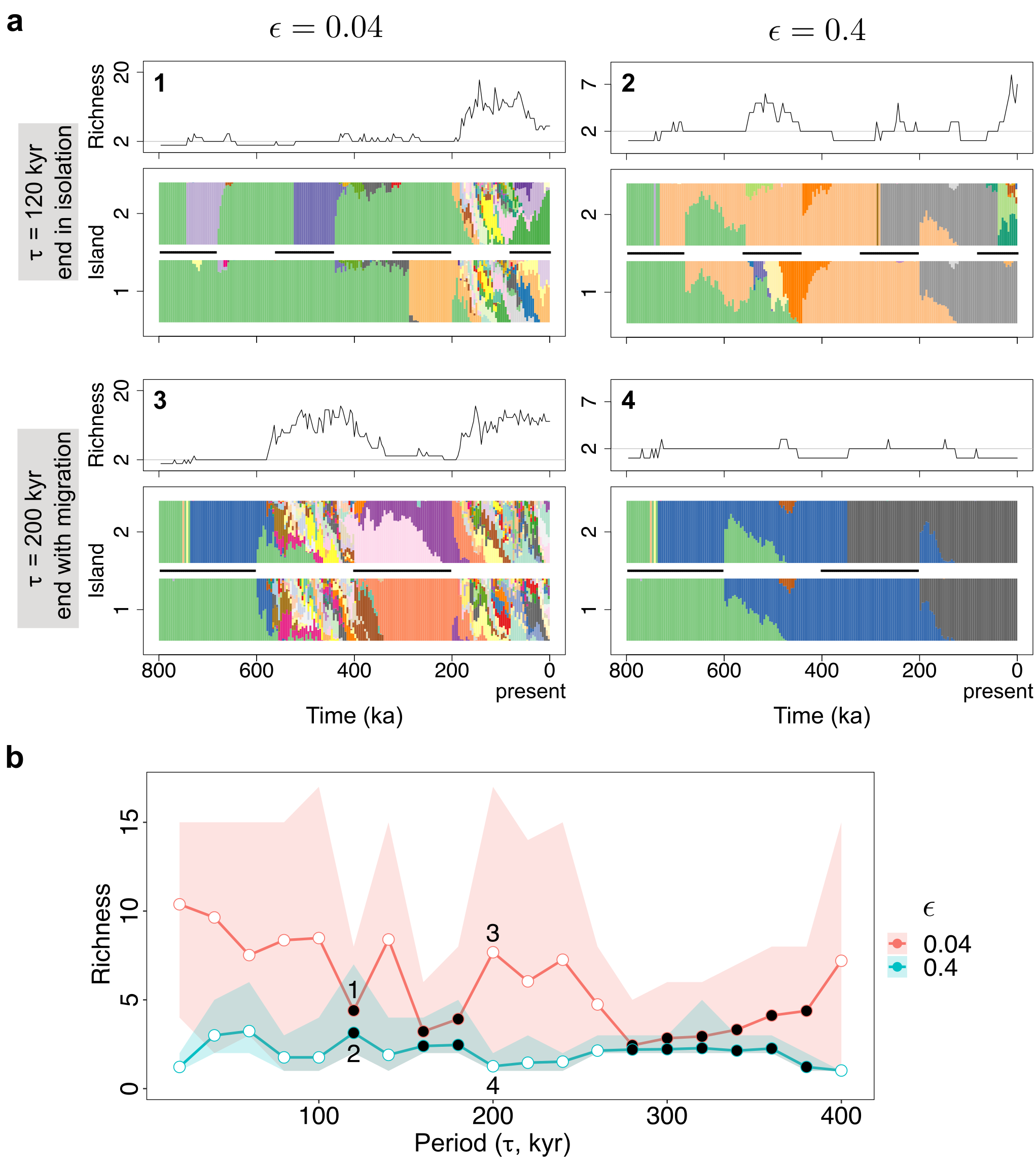}
\end{center}
\caption{ Speciation under periodic sea-level fluctuation for low ($\epsilon =0.04$) and high ($\epsilon =0.4$) migration probability.
(a) Richness trajectories and species abundances in single simulations of populations evolved with period equal to $\tau=120$ (1 and 2, end in isolation) and $\tau=200$ (3 and 4, end with migration) thousands of years. Note that the richness plots on the left (1, 3) have a different scale from the right (2, 4).
(b) Average species richness at the end of the simulations (present) over 50 independent realizations for varying $\tau$ (circles connected by trend lines).
The shadowed areas show a confidence interval of 90\%. White (black) circles indicate when the simulation ended with connected (isolated) islands; points 1 to 4 are exemplified in (a).}
 \label{fig:fig4}
\end{figure}

Simulations with regular cycles supported that a lower migration probability favored speciation in general (see species richness in Fig. \ref{fig:fig4}a and average richness in Fig. \ref{fig:fig4}b for $\epsilon = 0.04$ and $0.4$), but there was not such a clear trend in final richness in relation to cycle period (Fig. \ref{fig:fig4}b).
Still, there was a relative effect linking migration intensity and period, and whether the simulation ended with isolated or connected islands (Fig. \ref{fig:fig4}b): the periods that favored speciation with low migration ($\epsilon = 0.04$, red line) were opposed to those that favored speciation with high migration ($\epsilon = 0.4$, green line) and vice versa; moreover, shifts in the state of connection reversed this relative trend, implying that connection and isolation had opposite effects for low and high migration intensity. 
Figure \ref{fig:fig4}a exemplifies this inverted behavior. For $\epsilon=0.04$, speciation occurred during the connection phase; therefore, the final richness was higher when ending with migration. Whereas for $\epsilon=0.4$, most diversification occurred during the isolation period (more pronounced in Fig. \ref{fig:fig4}a-2), then richness was higher when ending in isolation. Interestingly, the observed increase in species richness during connection under low migration (Fig. \ref{fig:fig4}a-1,3) is contrary to that expected by the taxon pulse hypothesis \citep{Erwin1985,Halas2005}. 


\section*{Discussion}

We have explored a speciation model in a two-island system subjected to varying connectivity due to sea-level fluctuations. Individuals reproduce locally, constrained by genetic similarity, and populations evolve by neutral processes (genetic drift and mutation). 
We have shown that speciation (within islands) occurred due to episodic migration under conditions where it would not happen with strict isolation or continuous migration ($\epsilon \geq 0.04$, see Fig. \ref{fig:fig3})\footnote{For $B=2000$, speciation with continuous migration strictly occurs with $\epsilon < 0.03$, according to \cite{Princepe2022}.}.
Our results exhibited complex dynamic patterns in which species richness peaked at different times for different geographic settings (Fig. \ref{fig:fig2}) and were a byproduct of speciation pulses that depend on the migration probability and the isolation and connection times (Fig. \ref{fig:fig3} and \ref{fig:fig4}). Although transient population differentiation under variable migration is typically associated with strong selection \citep{Feder2019,Aguilee2011}, we were able to replicate this behavior in its absence, implying that neutral evolutionary processes can be relevant on short timescales, especially when dealing with populations out of equilibrium and dynamic landscapes.

Speciation was favored by low-intensity migrations in any frequency (Fig. \ref{fig:fig2} and \ref{fig:fig3}, left column) or by rare migration events with many migrants per event (Fig. \ref{fig:fig3}, right column), in agreement with previous models \citep{Yamaguchi-2013_first}. 
The heat plots in Figure \ref{fig:fig2} also illustrate this interdependence between migration intensity and duration of isolation and connection periods, where a region of single shared species formed in the crossing between shallow seabeds and high migration. We believe that speciation under these conditions may take longer times, following the trend of the heat maps. Furthermore, a higher richness was associated consistently with lower migration probability. 

Other neutral models of pulsed migration show, in general, a direct relationship between the average time to speciation and the time in isolation: lowest in isolation, intermediate under periodic migration, and highest under constant migration \citep{Linck2019}. 
Here, with the conditions shown in Figure \ref{fig:fig3}, $\epsilon=0.04$ and $0.4$, speciation did not occur with constant migration (Fig. \ref{fig:fig3}, top). However, divergence in allopatry took about 65 kyr (see the beginning of simulations in Fig. \ref{fig:fig4}a), whereas short instances of contact were sufficient to induce fast pulses of speciation (see Fig. \ref{fig:fig3}, $h=-100m$, for example).


The increase in the number of species beyond the total of two obtained in strict isolation (one in each island) seems to result from a combination of the speciation mechanisms investigated previously: founding populations, when the islands first exchange migrants of different species, and induced sympatric speciation, when species are shared, and divergence occurs faster \citep{Princepe2022}. 
\cite{He2019} also pointed out this potential for an exponential increase in richness due to the mixed dynamics of isolation with migration, suggesting a process for the evolution of biodiversity hotspots. Indeed, this complex dynamic presents interesting aspects, such as the coexistence of many species with gene flow (Fig. \ref{fig:fig4}a-1,3) and instances of speciation despite high migration (Fig. \ref{fig:fig4}a-2,4).

Pulsed speciation was a key finding in our model: species richness rapidly increased but returned to a few or single species per island after some time (see Fig. \ref{fig:fig3}, $h=-100m$, and Fig. \ref{fig:fig4}a-1,2,3). This dynamic effect would go unnoticed in a model that solely assesses populations in equilibrium. We observed two types of dynamics depending on the intensity of migration and regularity and time in connection/isolation: one that agrees with the taxon pulse hypothesis (speciation during the isolation phase, see Fig. \ref{fig:fig3}, $h=-100m$, and Fig. \ref{fig:fig4}a-2) and another with opposite behavior (speciation while connected, see Fig. \ref{fig:fig4}a-1,3)\citep{Erwin1985,Halas2005}. 
Our results revisit this hypothesis, highlighting a key aspect of the dynamics: speciation bursts can be ephemeral. With a long isolation period, the richness can revert to a single species on each island. Moreover, the timing of the geographic barriers can be asynchronous with the speciation times. These observations could explain the challenge of identifying the occurrence of diversification by taxon pulses \citep{Ober2003, Liebherr1990}. Indeed, extinctions were considered by Erwin as part of the pulses \citep{Erwin1985}, and are known to make it difficult to infer diversification rates \citep{Halas2005}. Factors not accounted for in our model, such as ecological interactions, selection, and reinforcement by secondary contact \citep{Aguilee2011}, can facilitate species persistence following the pulse.

The richness patterns that emerged in our model intertwine temporal and spatial dimensions, resembling the effect of intermittent barriers in nature, particularly those due to sea-level fluctuations.
Regions exhibiting high diversity and endemism show signatures of the eustatic fluctuations during the Pleistocene \citep{Hisheh1998,Hewitt2000,Lohman2011,Tscha2017}, such that vicariant processes alone cannot explain the genetic structure in certain taxa \citep{He2019, Cowie2006,Linck2019}.
In the Indo-Western Pacific, for instance, exposed land bridges putatively impacted terrestrial and aquatic species distributions \citep{Guo2015,Keyse2018,Lukoschek2007,Sholihah2021}. 
On the Brazilian coast, regions with broad and deep continental shelves sporadically facilitated population expansion, resulting in increased genetic exchange, lower differentiation, and reticulate evolution, suggesting diversification by taxon pulses \citep{baggio2017}.
Similar signatures persisted even when speciation occurred before the intermittent connections \citep{Lohman2011,Fauvelot2020,Esselstyn2009}. Populations admixture influenced intraspecific genetic and morphological variation \citep{Surridge1999, Ruedi1996}, and the presence of cryptic species implies hybridization zones and recent or ongoing speciation processes \citep{Fauvelot2020, Carpenter2011}.

Dissimilarities in richness patterns among taxa can be attributed to their distinct susceptibilities to varying connectivity and inherent characteristics of the genus, such as the speciation times. 
However, the most likely explanation is the spatial inhomogeneity of environmental conditions: climate change was the primary cause of sea-level fluctuations, exposing species to distinct ecological and environmental conditions \citep{Crandall2008, Heaney2005}. 
We acknowledge the limitations of our investigation, which focus on spatially assorting neutral genetic variation, but neglects potential selection gradients. Nonetheless, our model offers insights into the mechanisms of biological diversity assembly in dynamically connected systems with transient patterns.

Our work aligns with recent efforts emphasizing the significance of dynamic barriers and ephemeral effects to understanding complex environments \citep{Palamara2023, Reigada2015, Linck2019, He2019}, thereby emphasizing their relevance for conservation \citep{Randall1998, Rocha2007}. 
Further investigation to distinguish the indirect effects of varying migration, such as intra and interspecific genetic variability, hybridization, and reticulate evolution, would be a natural next step. Our study emphasizes the current need to integrate population genetics into island biogeography for predicting community data across different scales \citep{Overcast2023}.

\section*{Acknowledgment:}
This work was partly supported by the São Paulo Research Foundation (FAPESP), grants \#2018/11187-8 (DP), \#2019/20271-5 (MAMA), and \#2016/01343-7 and \#2021/14335-0 (ICTP-SAIFR). FMDM was supported by Coordenação de Aperfeiçoamento de Pessoal de Nível Superior -- Brazil (Finance Code 001). MAMA and SBLA were supported by Conselho Nacional de Pesquisas Científicas (CNPq), grants \#301082/2019-7 and \#311284/2021-3 respectively. SBLA acknowledge the computational support from Professor Carlos M. de Carvalho at LFTC-DFis-UFPR.

\section*{Conflict of interest: }
The authors declare that they have no conflict of interests.
{
\footnotesize
\bibliography{referencias}}

\begin{thebibliography}{10}
\expandafter\ifx\csname url\endcsname\relax
  \def\url#1{\texttt{#1}}\fi
\expandafter\ifx\csname urlprefix\endcsname\relax\def\urlprefix{URL }\fi
\providecommand{\bibinfo}[2]{#2}
\providecommand{\eprint}[2][]{\url{#2}}

\bibitem{Lewis2023}
\bibinfo{author}{Lewis, J.~A.}, \bibinfo{author}{Kandala, P.},
  \bibinfo{author}{Penley, M.~J.} \& \bibinfo{author}{Morran, L.~T.}
\newblock \bibinfo{title}{{Gene flow accelerates adaptation to a parasite}}.
\newblock \emph{\bibinfo{journal}{Evolution}} \bibinfo{pages}{qpad048}
  (\bibinfo{year}{2023}).

\bibitem{Matute2010}
\bibinfo{author}{Matute, D.~R.}
\newblock \bibinfo{title}{Reinforcement can overcome gene flow during
  speciation in drosophila}.
\newblock \emph{\bibinfo{journal}{Current Biology}}
  \textbf{\bibinfo{volume}{20}}, \bibinfo{pages}{2229--2233}
  (\bibinfo{year}{2010}).

\bibitem{Spurgin2014}
\bibinfo{author}{Spurgin, L.~G.}, \bibinfo{author}{Illera, J.~C.},
  \bibinfo{author}{Jorgensen, T.~H.}, \bibinfo{author}{Dawson, D.~A.} \&
  \bibinfo{author}{Richardson, D.~S.}
\newblock \bibinfo{title}{{Genetic and phenotypic divergence in an island bird:
  Isolation by distance, by colonization or by adaptation?}}
\newblock \emph{\bibinfo{journal}{Molecular Ecology}}
  \textbf{\bibinfo{volume}{23}}, \bibinfo{pages}{1028--1039}
  (\bibinfo{year}{2014}).

\bibitem{Claramunt2012}
\bibinfo{author}{Claramunt, S.} \emph{et~al.}
\newblock \bibinfo{title}{High dispersal ability inhibits speciation in a
  continental radiation of passerine birds}.
\newblock \emph{\bibinfo{journal}{Proceedings of the Royal Society B:
  Biological Sciences}} \textbf{\bibinfo{volume}{279}},
  \bibinfo{pages}{1567--1574} (\bibinfo{year}{2012}).

\bibitem{Lenormand2002}
\bibinfo{author}{Lenormand, T.}
\newblock \bibinfo{title}{{Gene flow and the limits to natural selection}}.
\newblock \emph{\bibinfo{journal}{Trends in Ecology {\&} Evolution}}
  \textbf{\bibinfo{volume}{17}}, \bibinfo{pages}{183--189}
  (\bibinfo{year}{2002}).

\bibitem{Yamaguchi-2013_first}
\bibinfo{author}{Yamaguchi, R.} \& \bibinfo{author}{Iwasa, Y.}
\newblock \bibinfo{title}{First passage time to allopatric speciation}.
\newblock \emph{\bibinfo{journal}{Interface Focus}}
  \textbf{\bibinfo{volume}{3}}, \bibinfo{pages}{20130026}
  (\bibinfo{year}{2013}).

\bibitem{Princepe2022}
\bibinfo{author}{Princepe, D.} \emph{et~al.}
\newblock \bibinfo{title}{{Diversity patterns and speciation processes in a
  two-island system with continuous migration}}.
\newblock \emph{\bibinfo{journal}{Evolution}} \textbf{\bibinfo{volume}{76}},
  \bibinfo{pages}{2260--2271} (\bibinfo{year}{2022}).

\bibitem{Garant2007}
\bibinfo{author}{Garant, D.}, \bibinfo{author}{Forde, S.~E.} \&
  \bibinfo{author}{Hendry, A.~P.}
\newblock \bibinfo{title}{{The multifarious effects of dispersal and gene flow
  on contemporary adaptation}}.
\newblock \emph{\bibinfo{journal}{Functional Ecology}}
  \textbf{\bibinfo{volume}{21}}, \bibinfo{pages}{434--443}
  (\bibinfo{year}{2007}).

\bibitem{Tigano2016}
\bibinfo{author}{Tigano, A.} \& \bibinfo{author}{Friesen, V.~L.}
\newblock \bibinfo{title}{{Genomics of local adaptation with gene flow}}.
\newblock \emph{\bibinfo{journal}{Molecular Ecology}}
  \textbf{\bibinfo{volume}{25}}, \bibinfo{pages}{2144--2164}
  (\bibinfo{year}{2016}).

\bibitem{coyne_speciation_2004}
\bibinfo{author}{Coyne, J.~A.} \& \bibinfo{author}{Orr, H.~A.}
\newblock \emph{\bibinfo{title}{Speciation}} (\bibinfo{publisher}{Sinauer
  Associates, Inc.}, \bibinfo{year}{2004}), \bibinfo{edition}{1} edn.

\bibitem{Everson2019}
\bibinfo{author}{Everson, K.~M.} \emph{et~al.}
\newblock \bibinfo{title}{{Speciation, gene flow, and seasonal migration in
  \textit{Catharus} thrushes (Aves:Turdidae)}}.
\newblock \emph{\bibinfo{journal}{Molecular Phylogenetics and Evolution}}
  \textbf{\bibinfo{volume}{139}}, \bibinfo{pages}{106564}
  (\bibinfo{year}{2019}).

\bibitem{Hoberg2008}
\bibinfo{author}{Hoberg, E.~P.} \& \bibinfo{author}{Brooks, D.~R.}
\newblock \bibinfo{title}{{A macroevolutionary mosaic: episodic host-switching,
  geographical colonization and diversification in complex host–parasite
  systems}}.
\newblock \emph{\bibinfo{journal}{Journal of Biogeography}}
  \textbf{\bibinfo{volume}{35}}, \bibinfo{pages}{1533--1550}
  (\bibinfo{year}{2008}).

\bibitem{Matthysen2005}
\bibinfo{author}{Matthysen, E.}
\newblock \bibinfo{title}{Density-dependent dispersal in birds and mammals}.
\newblock \emph{\bibinfo{journal}{Ecography}} \textbf{\bibinfo{volume}{28}},
  \bibinfo{pages}{403--416} (\bibinfo{year}{2005}).

\bibitem{White2010}
\bibinfo{author}{White, C.} \emph{et~al.}
\newblock \bibinfo{title}{Ocean currents help explain population genetic
  structure}.
\newblock \emph{\bibinfo{journal}{Proceedings of the Royal Society B:
  Biological Sciences}} \textbf{\bibinfo{volume}{277}},
  \bibinfo{pages}{1685--1694} (\bibinfo{year}{2010}).

\bibitem{Boizard2009}
\bibinfo{author}{Boizard, J.}, \bibinfo{author}{Magnan, P.} \&
  \bibinfo{author}{Angers, B.}
\newblock \bibinfo{title}{{Effects of dynamic landscape elements on fish
  dispersal: the example of creek chub (\textit{Semotilus atromaculatus})}}.
\newblock \emph{\bibinfo{journal}{Molecular Ecology}}
  \textbf{\bibinfo{volume}{18}}, \bibinfo{pages}{430--441}
  (\bibinfo{year}{2009}).

\bibitem{Tscha2017}
\bibinfo{author}{Tsch{\'{a}}, M.~K.} \emph{et~al.}
\newblock \bibinfo{title}{{Sea-level variations have influenced the demographic
  history of estuarine and freshwater fishes of the coastal plain of
  Paran{\'{a}}, Brazil}}.
\newblock \emph{\bibinfo{journal}{Journal of Fish Biology}}
  \textbf{\bibinfo{volume}{90}}, \bibinfo{pages}{968--979}
  (\bibinfo{year}{2017}).

\bibitem{Ludt2015}
\bibinfo{author}{Ludt, W.~B.} \& \bibinfo{author}{Rocha, L.~A.}
\newblock \bibinfo{title}{{Shifting seas: the impacts of Pleistocene sea-level
  fluctuations on the evolution of tropical marine taxa}}.
\newblock \emph{\bibinfo{journal}{Journal of Biogeography}}
  \textbf{\bibinfo{volume}{42}}, \bibinfo{pages}{25--38}
  (\bibinfo{year}{2015}).

\bibitem{Briggs2013}
\bibinfo{author}{Briggs, J.~C.} \& \bibinfo{author}{Bowen, B.~W.}
\newblock \bibinfo{title}{{Marine shelf habitat: biogeography and evolution}}.
\newblock \emph{\bibinfo{journal}{Journal of Biogeography}}
  \textbf{\bibinfo{volume}{40}}, \bibinfo{pages}{1023--1035}
  (\bibinfo{year}{2013}).

\bibitem{Hewitt2000}
\bibinfo{author}{Hewitt, G.}
\newblock \bibinfo{title}{{The genetic legacy of the Quaternary ice ages}}.
\newblock \emph{\bibinfo{journal}{Nature}} \textbf{\bibinfo{volume}{405}},
  \bibinfo{pages}{907--913} (\bibinfo{year}{2000}).

\bibitem{Guo2015}
\bibinfo{author}{Guo, Y.~Y.}, \bibinfo{author}{Luo, Y.~B.},
  \bibinfo{author}{Liu, Z.~J.} \& \bibinfo{author}{Wang, X.~Q.}
\newblock \bibinfo{title}{{Reticulate evolution and sea-level fluctuations
  together drove species diversification of slipper orchids
  (\textit{Paphiopedilum}) in South-East Asia}}.
\newblock \emph{\bibinfo{journal}{Molecular Ecology}}
  \textbf{\bibinfo{volume}{24}}, \bibinfo{pages}{2838--2855}
  (\bibinfo{year}{2015}).

\bibitem{baggio2017}
\bibinfo{author}{Baggio, R.~A.}, \bibinfo{author}{Stoiev, S.~B.},
  \bibinfo{author}{Spach, H.~L.} \& \bibinfo{author}{Boeger, W.~A.}
\newblock \bibinfo{title}{Opportunity and taxon pulse: the central influence of
  coastal geomorphology on genetic diversification and endemism of strict
  estuarine species}.
\newblock \emph{\bibinfo{journal}{Journal of Biogeography}}
  \textbf{\bibinfo{volume}{44}}, \bibinfo{pages}{1626--1639}
  (\bibinfo{year}{2017}).

\bibitem{Cowie2008}
\bibinfo{author}{Cowie, R.~H.} \& \bibinfo{author}{Holland, B.~S.}
\newblock \bibinfo{title}{{Molecular biogeography and diversification of the
  endemic terrestrial fauna of the Hawaiian Islands}}.
\newblock \emph{\bibinfo{journal}{Philosophical Transactions of the Royal
  Society B: Biological Sciences}} \textbf{\bibinfo{volume}{363}},
  \bibinfo{pages}{3363--3376} (\bibinfo{year}{2008}).

\bibitem{Yesson2009}
\bibinfo{author}{Yesson, C.}, \bibinfo{author}{Toomey, N.~H.} \&
  \bibinfo{author}{Culham, A.}
\newblock \bibinfo{title}{{\textit{Cyclamen}: Time, sea and speciation
  biogeography using a temporally calibrated phylogeny}}.
\newblock \emph{\bibinfo{journal}{Journal of Biogeography}}
  \textbf{\bibinfo{volume}{36}}, \bibinfo{pages}{1234--1252}
  (\bibinfo{year}{2009}).

\bibitem{Fauvelot2020}
\bibinfo{author}{Fauvelot, C.} \emph{et~al.}
\newblock \bibinfo{title}{{Phylogeographical patterns and a cryptic species
  provide new insights into Western Indian Ocean giant clams phylogenetic
  relationships and colonization history}}.
\newblock \emph{\bibinfo{journal}{Journal of Biogeography}}
  \textbf{\bibinfo{volume}{47}}, \bibinfo{pages}{1086--1105}
  (\bibinfo{year}{2020}).

\bibitem{Erwin1979}
\bibinfo{author}{Erwin, T.~L.}
\newblock \bibinfo{title}{{Thoughts on the Evolutionary History of Ground
  Beetles: Hypotheses Generated from Comparative Faunal Analyses of Lowland
  Forest Sites in Temperate and Tropical Regions}}.
\newblock In \bibinfo{editor}{{Erwin, T.L., Ball, G.E., Whitehead, D.R.,
  Halpern}, A.} (ed.) \emph{\bibinfo{booktitle}{Carabid Beetles: Their
  Evolution, Natural History, and Classification}}, \bibinfo{pages}{539--592}
  (\bibinfo{publisher}{Springer Netherlands}, \bibinfo{address}{Dordrecht},
  \bibinfo{year}{1979}).

\bibitem{Erwin1985}
\bibinfo{author}{Erwin, T.~L.}
\newblock \bibinfo{title}{{The taxon pulse: A general pattern of lineage
  radiation and extinction among carabid beetles}}.
\newblock In \bibinfo{editor}{Ball, G.~E.} (ed.)
  \emph{\bibinfo{booktitle}{Taxonomy, phylogeny, and zoogeography of beetles
  and ant: A volume dedicated to the memory of Philip Jackson Darlington Jr.}},
  \bibinfo{pages}{437--472} (\bibinfo{publisher}{Dr. W. Junk b.v. Publishers},
  \bibinfo{address}{The Hague}, \bibinfo{year}{1985}).

\bibitem{Halas2005}
\bibinfo{author}{Halas, D.}, \bibinfo{author}{Zamparo, D.} \&
  \bibinfo{author}{Brooks, D.~R.}
\newblock \bibinfo{title}{{A historical biogeographical protocol for studying
  biotic diversification by taxon pulses}}.
\newblock \emph{\bibinfo{journal}{{Journal of Biogeography}}}
  \textbf{\bibinfo{volume}{32}}, \bibinfo{pages}{249--260}
  (\bibinfo{year}{2005}).

\bibitem{Lim2008}
\bibinfo{author}{Lim, B.~K.}
\newblock \bibinfo{title}{{Historical biogeography of New World emballonurid
  bats (tribe Diclidurini): taxon pulse diversification}}.
\newblock \emph{\bibinfo{journal}{Journal of Biogeography}}
  \textbf{\bibinfo{volume}{35}}, \bibinfo{pages}{1385--1401}
  (\bibinfo{year}{2008}).

\bibitem{Hoberg2010}
\bibinfo{author}{Hoberg, E.~P.} \& \bibinfo{author}{Brooks, D.}
\newblock \bibinfo{title}{{Beyond vicariance: integrating taxon pulses,
  ecological fitting and oscillation in evolution and historical
  biogeography}}.
\newblock In \bibinfo{editor}{{S Morand}} \& \bibinfo{editor}{{B Krasnov}}
  (eds.) \emph{\bibinfo{booktitle}{The Biogeography of Host-Parasite
  Interactions}}, \bibinfo{pages}{7--20} (\bibinfo{publisher}{Oxford University
  Press}, \bibinfo{address}{Oxford}, \bibinfo{year}{2010}),
  \bibinfo{edition}{1} edn.

\bibitem{Schweizer2010}
\bibinfo{author}{Schweizer, M.}, \bibinfo{author}{Seehausen, O.},
  \bibinfo{author}{G{\"{u}}ntert, M.} \& \bibinfo{author}{Hertwig, S.~T.}
\newblock \bibinfo{title}{{The evolutionary diversification of parrots supports
  a taxon pulse model with multiple trans-oceanic dispersal events and local
  radiations}}.
\newblock \emph{\bibinfo{journal}{Molecular Phylogenetics and Evolution}}
  \textbf{\bibinfo{volume}{54}}, \bibinfo{pages}{984--994}
  (\bibinfo{year}{2010}).

\bibitem{Bouchard2005}
\bibinfo{author}{Bouchard, P.}, \bibinfo{author}{Brooks, D.~R.} \&
  \bibinfo{author}{Yeates, D.~K.}
\newblock \bibinfo{title}{{Mosaic macroevolution in Australian Wet Tropics
  arthropods: community assemblage by taxon pulses.}}
\newblock In \emph{\bibinfo{booktitle}{Tropical rainforests: past, present and
  future}}, \bibinfo{pages}{425--469} (\bibinfo{publisher}{University of
  Chicago Press}, \bibinfo{address}{Chicago}, \bibinfo{year}{2005}).

\bibitem{Esseghir2000}
\bibinfo{author}{Esseghir, S.}, \bibinfo{author}{Ready, P.~D.} \&
  \bibinfo{author}{Ben-Ismail, R.}
\newblock \bibinfo{title}{{Speciation of \textit{Phlebotomus} sandflies of the
  subgenus \textit{Larroussius} coincided with the late Miocene-Pliocene
  aridification of the Mediterranean subregion}}.
\newblock \emph{\bibinfo{journal}{Biological Journal of the Linnean Society}}
  \textbf{\bibinfo{volume}{70}}, \bibinfo{pages}{189--219}
  (\bibinfo{year}{2000}).

\bibitem{He2019}
\bibinfo{author}{He, Z.} \emph{et~al.}
\newblock \bibinfo{title}{{Speciation with gene flow via cycles of isolation
  and migration: insights from multiple mangrove taxa}}.
\newblock \emph{\bibinfo{journal}{National Science Review}}
  \textbf{\bibinfo{volume}{6}}, \bibinfo{pages}{275--288}
  (\bibinfo{year}{2019}).

\bibitem{Linck2019}
\bibinfo{author}{Linck, E.} \& \bibinfo{author}{Battey, C.~J.}
\newblock \bibinfo{title}{{On the relative ease of speciation with periodic
  gene flow}}.
\newblock \emph{\bibinfo{journal}{bioRxiv}} \bibinfo{pages}{758664}
  (\bibinfo{year}{2019}).

\bibitem{Albert2017}
\bibinfo{author}{Albert, J.~S.}, \bibinfo{author}{Schoolmaster, D.~R.},
  \bibinfo{author}{Tagliacollo, V.} \& \bibinfo{author}{Duke-Sylvester, S.~M.}
\newblock \bibinfo{title}{{Barrier Displacement on a Neutral Landscape: Toward
  a Theory of Continental Biogeography}}.
\newblock \emph{\bibinfo{journal}{Systematic Biology}}
  \textbf{\bibinfo{volume}{66}}, \bibinfo{pages}{167--182}
  (\bibinfo{year}{2017}).

\bibitem{Rice2009}
\bibinfo{author}{Rice, S.~H.} \& \bibinfo{author}{Papadopoulos, A.}
\newblock \bibinfo{title}{{Evolution with Stochastic Fitness and Stochastic
  Migration}}.
\newblock \emph{\bibinfo{journal}{PLOS ONE}} \textbf{\bibinfo{volume}{4}},
  \bibinfo{pages}{e7130} (\bibinfo{year}{2009}).

\bibitem{Aguilee2011}
\bibinfo{author}{Aguil{\'{e}}e, R.}, \bibinfo{author}{Lambert, A.} \&
  \bibinfo{author}{Claessen, D.}
\newblock \bibinfo{title}{{Ecological speciation in dynamic landscapes}}.
\newblock \emph{\bibinfo{journal}{Journal of Evolutionary Biology}}
  \textbf{\bibinfo{volume}{24}}, \bibinfo{pages}{2663--2677}
  (\bibinfo{year}{2011}).

\bibitem{Peniston2019}
\bibinfo{author}{Peniston, J.~H.}, \bibinfo{author}{Barfield, M.} \&
  \bibinfo{author}{Holt, R.~D.}
\newblock \bibinfo{title}{{Pulsed Immigration Events Can Facilitate Adaptation
  to Harsh Sink Environments}}.
\newblock \emph{\bibinfo{journal}{The American Naturalist}}
  \textbf{\bibinfo{volume}{194}}, \bibinfo{pages}{316--333}
  (\bibinfo{year}{2019}).

\bibitem{Aubree2023}
\bibinfo{author}{Aubree, F.}, \bibinfo{author}{Lac, B.},
  \bibinfo{author}{Mailleret, L.} \& \bibinfo{author}{Calcagno, V.}
\newblock \bibinfo{title}{{Migration pulsedness alters patterns of allele
  fixation and local adaptation in a mainland-island model}}.
\newblock \emph{\bibinfo{journal}{Evolution}} \textbf{\bibinfo{volume}{77}},
  \bibinfo{pages}{718--730} (\bibinfo{year}{2023}).

\bibitem{Palamara2023}
\bibinfo{author}{Palamara, G.~M.} \emph{et~al.}
\newblock \bibinfo{title}{Biodiversity dynamics in landscapes with fluctuating
  connectivity}.
\newblock \emph{\bibinfo{journal}{Ecography}} \bibinfo{pages}{e06385}
  (\bibinfo{year}{2023}).

\bibitem{Reigada2015}
\bibinfo{author}{Reigada, C.}, \bibinfo{author}{Schreiber, S.~J.},
  \bibinfo{author}{Altermatt, F.} \& \bibinfo{author}{Holyoak, M.}
\newblock \bibinfo{title}{Metapopulation dynamics on ephemeral patches}.
\newblock \emph{\bibinfo{journal}{American Naturalist}}
  \textbf{\bibinfo{volume}{185}}, \bibinfo{pages}{183--195}
  (\bibinfo{year}{2015}).

\bibitem{Feder2019}
\bibinfo{author}{Feder, A.~F.}, \bibinfo{author}{Pennings, P.~S.},
  \bibinfo{author}{Hermisson, J.} \& \bibinfo{author}{Petrov, D.~A.}
\newblock \bibinfo{title}{{Evolutionary Dynamics in Structured Populations
  Under Strong Population Genetic Forces}}.
\newblock \emph{\bibinfo{journal}{G3 Genes|Genomes|Genetics}}
  \textbf{\bibinfo{volume}{9}}, \bibinfo{pages}{3395--3407}
  (\bibinfo{year}{2019}).

\bibitem{MacArthur1967}
\bibinfo{author}{MacArthur, R.~H.} \& \bibinfo{author}{Wilson, E.~O.}
\newblock \emph{\bibinfo{title}{{The Theory of Island Biogeography}}}
  (\bibinfo{publisher}{Princeton University Press},
  \bibinfo{address}{Princeton}, \bibinfo{year}{1967}).

\bibitem{de2009global}
\bibinfo{author}{de~Aguiar, M. A.~M.}, \bibinfo{author}{Baranger, M.},
  \bibinfo{author}{Baptestini, E.}, \bibinfo{author}{Kaufman, L.} \&
  \bibinfo{author}{Bar-Yam, Y.}
\newblock \bibinfo{title}{Global patterns of speciation and diversity}.
\newblock \emph{\bibinfo{journal}{Nature}} \textbf{\bibinfo{volume}{460}},
  \bibinfo{pages}{384--387} (\bibinfo{year}{2009}).

\bibitem{Manzo_Peliti_1994}
\bibinfo{author}{Manzo, F.} \& \bibinfo{author}{Peliti, L.}
\newblock \bibinfo{title}{Geographic speciation in the {Derrida-Higgs} model of
  species formation}.
\newblock \emph{\bibinfo{journal}{Journal of Physics A: Mathematical and
  General}} \textbf{\bibinfo{volume}{27}}, \bibinfo{pages}{7079}
  (\bibinfo{year}{1994}).

\bibitem{Spratt2016}
\bibinfo{author}{Spratt, R.~M.} \& \bibinfo{author}{Lisiecki, L.~E.}
\newblock \bibinfo{title}{{A Late Pleistocene sea level stack}}.
\newblock \emph{\bibinfo{journal}{Clim. Past}} \textbf{\bibinfo{volume}{12}},
  \bibinfo{pages}{1079--1092} (\bibinfo{year}{2016}).

\bibitem{higgs_stochastic_1991}
\bibinfo{author}{Higgs, P.~G.} \& \bibinfo{author}{Derrida, B.}
\newblock \bibinfo{title}{Stochastic models for species formation in evolving
  populations}.
\newblock \emph{\bibinfo{journal}{Journal of Physics A: Mathematical and
  General}} \textbf{\bibinfo{volume}{24}}, \bibinfo{pages}{L985--L991}
  (\bibinfo{year}{1991}).

\bibitem{aguiar_speciation_2017}
\bibinfo{author}{de~Aguiar, M. A.~M.}
\newblock \bibinfo{title}{{Speciation in the Derrida-Higgs model with finite
  genomes and spatial populations}}.
\newblock \emph{\bibinfo{journal}{Journal of Physics A: Mathematical and
  Theoretical}} \textbf{\bibinfo{volume}{50}}, \bibinfo{pages}{085602}
  (\bibinfo{year}{2017}).

\bibitem{Ober2003}
\bibinfo{author}{Ober, K.~A.}
\newblock \bibinfo{title}{{Arboreality and morphological evolution in ground
  beetles (Carabidae: Harpalinae): testing the taxon pulse model}}.
\newblock \emph{\bibinfo{journal}{Evolution}} \textbf{\bibinfo{volume}{57}},
  \bibinfo{pages}{1343--1358} (\bibinfo{year}{2003}).

\bibitem{Liebherr1990}
\bibinfo{author}{Liebherr, J.~K.} \& \bibinfo{author}{Hajek, A.~E.}
\newblock \bibinfo{title}{{A cladistic test of taxon cycle and taxon pulse
  hypotheses}}.
\newblock \emph{\bibinfo{journal}{Cladistics}} \textbf{\bibinfo{volume}{6}},
  \bibinfo{pages}{39--59} (\bibinfo{year}{1990}).

\bibitem{Hisheh1998}
\bibinfo{author}{Hisheh, S.}, \bibinfo{author}{Westerman, M.} \&
  \bibinfo{author}{Schmitt, L.~H.}
\newblock \bibinfo{title}{{Biogeography of the Indonesian archipelago:
  mitochondrial DNA variation in the fruit bat, \textit{Eonycteris spelaea}}}.
\newblock \emph{\bibinfo{journal}{Biological Journal of the Linnean Society}}
  \textbf{\bibinfo{volume}{65}}, \bibinfo{pages}{329--345}
  (\bibinfo{year}{1998}).

\bibitem{Lohman2011}
\bibinfo{author}{Lohman, D.~J.} \emph{et~al.}
\newblock \bibinfo{title}{{Biogeography of the Indo-Australian Archipelago}}.
\newblock \emph{\bibinfo{journal}{Annual Review of Ecology, Evolution, and
  Systematics}} \textbf{\bibinfo{volume}{42}}, \bibinfo{pages}{205--226}
  (\bibinfo{year}{2011}).

\bibitem{Cowie2006}
\bibinfo{author}{Cowie, R.~H.} \& \bibinfo{author}{Holland, B.~S.}
\newblock \bibinfo{title}{{Dispersal is fundamental to biogeography and the
  evolution of biodiversity on oceanic islands}}.
\newblock \emph{\bibinfo{journal}{Journal of Biogeography}}
  \textbf{\bibinfo{volume}{33}}, \bibinfo{pages}{193--198}
  (\bibinfo{year}{2006}).

\bibitem{Keyse2018}
\bibinfo{author}{Keyse, J.} \emph{et~al.}
\newblock \bibinfo{title}{{Historical divergences associated with intermittent
  land bridges overshadow isolation by larval dispersal in co-distributed
  species of \textit{Tridacna} giant clams}}.
\newblock \emph{\bibinfo{journal}{Journal of Biogeography}}
  \textbf{\bibinfo{volume}{45}}, \bibinfo{pages}{848--858}
  (\bibinfo{year}{2018}).

\bibitem{Lukoschek2007}
\bibinfo{author}{Lukoschek, V.}, \bibinfo{author}{Waycott, M.} \&
  \bibinfo{author}{Marsh, H.}
\newblock \bibinfo{title}{{Phylogeography of the olive sea snake,
  \textit{Aipysurus laevis} (Hydrophiinae) indicates Pleistocene range
  expansion around northern Australia but low contemporary gene flow}}.
\newblock \emph{\bibinfo{journal}{Molecular Ecology}}
  \textbf{\bibinfo{volume}{16}}, \bibinfo{pages}{3406--3422}
  (\bibinfo{year}{2007}).

\bibitem{Sholihah2021}
\bibinfo{author}{Sholihah, A.} \emph{et~al.}
\newblock \bibinfo{title}{{Impact of Pleistocene Eustatic Fluctuations on
  Evolutionary Dynamics in Southeast Asian Biodiversity Hotspots}}.
\newblock \emph{\bibinfo{journal}{Systematic Biology}}
  \textbf{\bibinfo{volume}{70}}, \bibinfo{pages}{940--960}
  (\bibinfo{year}{2021}).

\bibitem{Esselstyn2009}
\bibinfo{author}{Esselstyn, J.~A.}, \bibinfo{author}{Timm, R.~M.} \&
  \bibinfo{author}{Brown, R.~M.}
\newblock \bibinfo{title}{{Do geological or climatic processes drive speciation in dynamic archipelagos? The tempo and mode of diversification in Southeast Asian shrews}}.
\newblock \emph{\bibinfo{journal}{Evolution}} \textbf{\bibinfo{volume}{63}},
  \bibinfo{pages}{2595--2610} (\bibinfo{year}{2009}).

\bibitem{Surridge1999}
\bibinfo{author}{Surridge, A.~K.}, \bibinfo{author}{Timmins, R.~J.},
  \bibinfo{author}{Hewitt, G.~M.} \& \bibinfo{author}{Bell, D.~J.}
\newblock \bibinfo{title}{{Striped rabbits in Southeast Asia}}.
\newblock \emph{\bibinfo{journal}{Nature}} \textbf{\bibinfo{volume}{400}},
  \bibinfo{pages}{726--726} (\bibinfo{year}{1999}).

\bibitem{Ruedi1996}
\bibinfo{author}{Ruedi, M.} \& \bibinfo{author}{Fumagalli, L.}
\newblock \bibinfo{title}{{Genetic structure of Gymnures (genus
  \textit{Hylomys}; Erinaceidae) on continental islands of Southeast Asia:
  historical effects of fragmentation}}.
\newblock \emph{\bibinfo{journal}{Journal of Zoological Systematics and
  Evolutionary Research}} \textbf{\bibinfo{volume}{34}},
  \bibinfo{pages}{153--162} (\bibinfo{year}{1996}).

\bibitem{Carpenter2011}
\bibinfo{author}{Carpenter, K.~E.} \emph{et~al.}
\newblock \bibinfo{title}{{Comparative phylogeography of the coral triangle and
  implications for marine management}}.
\newblock \emph{\bibinfo{journal}{Journal of Marine Biology}}
  \textbf{\bibinfo{volume}{2011}} (\bibinfo{year}{2011}).

\bibitem{Crandall2008}
\bibinfo{author}{Crandall, E.~D.} \emph{et~al.}
\newblock \bibinfo{title}{{Comparative phylogeography of two seastars and their
  ectosymbionts within the Coral Triangle}}.
\newblock \emph{\bibinfo{journal}{Molecular Ecology}}
  \textbf{\bibinfo{volume}{17}}, \bibinfo{pages}{5276--5290}
  (\bibinfo{year}{2008}).

\bibitem{Heaney2005}
\bibinfo{author}{Heaney, L.~R.}, \bibinfo{author}{Walsh, J.~S.} \&
  \bibinfo{author}{Peterson, A.~T.}
\newblock \bibinfo{title}{{The roles of geological history and colonization
  abilities in genetic differentiation between mammalian populations in the
  Philippine archipelago}}.
\newblock \emph{\bibinfo{journal}{Journal of Biogeography}}
  \textbf{\bibinfo{volume}{32}}, \bibinfo{pages}{229--247}
  (\bibinfo{year}{2005}).

\bibitem{Randall1998}
\bibinfo{author}{Randall, J.~E.}
\newblock \bibinfo{title}{{Zoogeography of shore fishes of the Indo-Pacific
  region}}.
\newblock \emph{\bibinfo{journal}{Zoological Studies}}
  \textbf{\bibinfo{volume}{37}}, \bibinfo{pages}{227--268}
  (\bibinfo{year}{1998}).

\bibitem{Rocha2007}
\bibinfo{author}{Rocha, L.~A.}, \bibinfo{author}{Craig, M.~T.} \&
  \bibinfo{author}{Bowen, B.~W.}
\newblock \bibinfo{title}{{Phylogeography and the conservation of coral reef
  fishes}}.
\newblock \emph{\bibinfo{journal}{Coral Reefs}} \textbf{\bibinfo{volume}{26}},
  \bibinfo{pages}{501--512} (\bibinfo{year}{2007}).

\bibitem{Overcast2023}
\bibinfo{author}{Overcast, I.} \emph{et~al.}
\newblock \bibinfo{title}{Towards a genetic theory of island biogeography:
  Inferring processes from multidimensional community-scale data}.
\newblock \emph{\bibinfo{journal}{Global Ecology and Biogeography}}
  \textbf{\bibinfo{volume}{32}}, \bibinfo{pages}{4--23} (\bibinfo{year}{2023}).

\end{thebibliography}

\end{document}